# Control of dissipation in superconducting films by magnetic stray fields

A. Gomez[1], D. A. Gilbert[2], E. M. Gonzalez[1,3], Kai Liu[2] and J. L. Vicent[1,3]

[1] Departamento de Física de Materiales, Facultad de Ciencias Fisicas, Universidad Complutense, 28040 Madrid, Spain

[2] Department of Physics, University of California, Davis, CA 95616, USA

[3] IMDEA-Nanociencia, Cantoblanco, 28049 Madrid, Spain

## Abstract

Hybrid superconducting/magnetic nanostructures on Si substrates have been built with identical physical dimensions but different magnetic configurations. By constructing arrays based on Co-dots with in-plane, out-of-plane, and vortex state magnetic configurations, the stray fields are systematically tuned. Dissipation in the mixed state of superconductors can be decreased (increased) by several orders of magnitude by decreasing (increasing) the stray magnetic fields. Furthermore, ordering of the stray fields over the entire array helps to suppress dissipation and enhance commensurability effects increasing the number of dissipation minima.

Superconductivity and magnetism are generally considered competing effects, but these two long range order phenomena, with proper system design, have been shown to develop cooperative behavior, enhancing the system's properties and giving rise to new and remarkable phenomena. Prior studies on superconductor/ferromagnetic hybrid systems have shown that magnetic structures can strongly influence the nucleation of superconductivity (1), as well as enhance pinning of the superconducting vortex lattice (2). Aladyshkin et al. (1), Perez et al. (3), and Van Bael et al. (4, 5) have emphasized the role stray field plays in determining the behavior of the superconductor. Furthermore, periodic arrays of nanoelements embedded in superconductors have been shown to significantly alter their intrinsic properties, for instance, changing the dynamic phases of the vortex matter (6,7) or vortex channeling and commensurability effects (8 -13). Recently, it has been shown that by switching to low flux flow dissipation in a regime with intermediate pinning strength, the stability of the superconducting state can be promoted (14).

All these phenomena offer the opportunity to enhance the performance of superconducting devices by controlling the dissipation induced by the movement of vortices. In this letter, we present work focused on controlling the mixed state dissipation from superconducting vortices using arrays of magnetic nanostructures as pinning sites, specifically investigating the role of the magnetic stray fields. By tailoring the magnetic structure of buried Co dots with identical physical dimensions (including dots with out-of-plane, in-plane, and vortex state magnetic configurations) the stray fields are systematically tuned and configuration dependent pinning is investigated. This approach eliminates the usual complications from structural variations at the superconductor / ferromagnet interfaces. Magnetoresistance measurements are used to determine the convoluted effects of stray fields and the periodic array. In contrast to previous studies (15), which have shown that an increase in the magnetic stray fields produces an increase in the critical current, we study the vortex dynamics beyond the critical current. We show that, in this regime where vortices are already moving, an increase in the magnetic stray field generates an increase in dissipation. These results systematically probe the main role of stray fields in superconducting vortex dynamics.

Arrays based on circular Co/Pd nanodots (200 nm diameter and 42 nm thickness) were fabricated using electron beam lithography, in conjunction with magnetron sputtering in an ultrahigh vacuum chamber with a base pressure of $1x10^{-8}$ Torr. The nanodots are polycrystalline, arranged on a rectangular lattice (400 nm x 600 nm spacing) covering an area of 100 x 100 $\mu m^2$. The structure of the Co and Pd layers are chosen so that different remanent magnetic states are obtained. Three arrays have been fabricated: a) $[Pd(0.6nm)/Co(0.4nm)]_{40}$ multilayer, b) Pd(5nm)/Co(35nm) bilayer, and c) Pd(24nm)/Co(16nm) bilayer; a 2 nm Pd capping layer was deposited on top to prevent oxidation. Finally, a 100 nm thick Nb film was deposited by magnetron sputtering on top of the arrays. Standard photolithography and ion etching techniques were used to define a cross-shaped, 40 µm wide bridge centered on the array, forming the magnetic / superconductor hybrid structure.

The sample magnetic characterization and transport measurement techniques are the same as that reported by Perez et al (16). Briefly, the following experimental procedures are used to set the nanodot remanent magnetic state: i) the ac-demagnetized state was realized by applying a decreasing ac magnetic field; ii) the dc-demagnetized state (saturation remanent state) is induced in the sample by applying a 20 kOe saturating magnetic field and then switching it off. Magnetic characterization was performed by magnetic force microscopy (MFM) and magneto-optical Kerr effect (MOKE) on identical nanomagnet arrays without the Nb layer. Magnetometry and the first-order reversal curve method (FORC) (17, 18) were employed to determine and realize the following magnetic states: a) out-of plane magnetization (OP) in $[Pd(0.6nm)/Co(0.4nm)]_{40}$ after dc-demagnetization, b) remanent magnetic vortex state (VS) in Pd(5nm)/Co(35nm) after in-plane saturation, and c) in-plane single domain (IPSD) state in Pd(24nm)/Co(16nm) after in-plane saturation

For nanodots in array a), the Co/Pd multilayers exhibit perpendicular anisotropy; the sputtering pressure was tuned to 12 mTorr to realize single domain state at remanence after saturation (19), as illustrated in Fig. 2 inset for a reference thin film sample. The VS and IPSD states in Co/Pd bilayer nanodots are illustrated in insets of Figs. 3 and 4, for array b) and c) respectively, consistent with prior studies on 200nm diameter Co nanodisks embedded in nanowires (20). Nanodots of array b) exhibit highly pinched hysteresis loops with minimum remanence (Fig. 3 inset). The corresponding FORC distribution (not shown) has distinct features related to vortex nucleation/annihilation fields. Nanodots of array c) are characteristic of single domain reversal, similar to those observed earlier (18).

Using the OOMMF simulation (21), stray fields in each magnetic state have been calculated. After the Nb deposition and patterning, a commercial helium cryostat with variable temperature insert and a superconducting solenoid is used for the magneto-transport measurements. Small magnetic fields perpendicular applied to the sample plane are used for these measurements, which do not change the remanent magnetic states.

Figure 1 shows a comparison of the dissipation data at T=0.985 $T_c$ for the three stray field configurations which correspond to the OP dc-demagnetized configuration (triangles in blue), IPSD configuration (squares in red), and VS configuration (dots in green). The OOMMF simulated stray fields are shown in Fig. 1 insets, illustrating that the OP case has the largest stray fields, the flux-closure VS has the least stray field, and the IPSD case is in between. Even though pinning strength has been shown to be higher for higher stray field (15), the present experimental results show that, when vortices are moving, an increase in the stray field leads to an increase in the dissipation. That is, the OP array yields the largest dissipation, the VS sample produces the smallest dissipation, and the dissipation value of the IPSD sample is in between.

More interesting physical insights can be extracted by examining the periodic minima dissipation. These minima are induced by commensurability effects between the vortex lattice and the nanodot array; see Ref. (2) and references therein. In the case of magnetoresistance minima induced by OP dc-demagnetized dots, the periodic minima distribution exhibits a clear asymmetry in that there are more minima for positive applied magnetic fields than those for negative ones. This effect is well understood since the pinning force can be either attractive or repulsive depending on the relative alignment (parallel or antiparallel) between the superconducting vortices and the stray field generated by the magnetic nanodot (22). Furthermore, for the VS sample the magnetoresistance data show new commensurability effects with additional minima (occurring in-between the large, sharp minima) which are generated by fractional matching fields (23). Another relevant result is that the monotonous background dissipation can diminish more than two orders of magnitude as the stray field is reduced from OP to VS. Thus, stray field configurations play a leading role in the mechanisms that govern both contributions to dissipation: sharp minima in dissipation (induced by matching effect) and the usual monotonous dissipation (background).

Further results are obtained by tuning, case by case, the different stray fields using the suitable array configuration. Fig. 2 shows the experimental magnetoresistance data for the case of stray fields generated by the OP sample. Results are shown for three cases: i) ac-demagnetized state (dots in blue), ii) positively dc-demagnetized state (magnetization remains parallel to positive magnetic field direction) (triangles in red) and iii) negatively dc-demagnetized state (magnetization parallel to negative magnetic field) (triangles in green). The measurements show that the lowest dissipation corresponds to the ac-demagnetized state with random up and down domains. In contrast, the dissipation increases by more than an order of magnitude when the entire array remains saturated in a particular orientation.  Comparing the positively (ii) and negatively (iii) saturated states the peak asymmetry is shown to switch sides, consistent with the explanation given above.

In Fig. 3, experimental results obtained for the VS sample are shown. We compare the magnetoresistance obtained for the disordered state with random polarity (where the magnetic vortex cores are randomly oriented up or down) with the ordered case with aligned polarity where all the cores are pointing in the positive direction out of plane (this state is achieved by an out of plane field of +20 kOe and switching it off) . In contrast to the trend seen in OP sample in Fig. 2, the random polarity configuration shows slightly larger dissipation compared to the aligned polarity case.

We lastly investigate the behavior of the IPSD sample, shown in Fig. 4, where each dot is in an in-plane single domain state.  In this case, disordered state corresponds to the case where the magnetization direction varies randomly from one dot to another. By contrast, after applying +20 kOe in the plane of the film along the short side of the unit cell and switching it off, an ordered state is obtained where all the magnetizations are pointing in the same direction in the dot plane. Similarly to the VS sample, the ordered state (triangles) shows a decrease in the background relative to the disordered state (dots).  In addition, extra minima appear in the ordered state, showing an increase in the commensurability effect.

From the experimental results obtained for the VS and IPSD samples, and taking into account that the *local* stray field generated by each dot is the same for both ordered and disordered states; a straightforward and consistent picture arises. By ordering the local magnetic stray fields created by the nanodots, an ordered magnetic landscape is created, producing an enhancement of the superconducting vortex lattice pinning. This influences the vortex lattice dynamics: First, it strongly reduces the usual monotonous background dissipation in comparison with the dissipation induced by the random distribution. Second, it enhances the commensurability effects and new minima show up.

In summary, vortex dynamics in superconducting films was controlled by tailoring the stray fields produced by buried magnetic nanodots. Choosing the appropriate magnetic configuration, these stray fields enhance or weaken the two types of dissipations which are found in these hybrid systems: i) the monotonous background dissipation and ii) the sharp and periodic dissipation minima. On one hand, the background dissipation can be enhanced up to two orders of magnitude by increasing the magnetic stray field. On the other hand, ordering the magnetic stray field can induce a decrease in dissipation and an increase in the number of matching minima due to the pinning of the vortex lattice. These results demonstrate a technique to tailor the superconducting dissipation using magnetic nanodots by tuning the stray fields.

This work has been supported by Spanish MINECO, FIS2008-06249 (Grupo Consolidado), Consolider CSD2007-00010 and CAM S2009/MAT-1726. Work at UCD has been supported by the US NSF (DMR-1008791 and ECCS-0925626).

**References**

1. A. Yu. Aladyshkin, A. V. Silhanek, W. Gillijns, and V. V. Moshchalkov, Supercond. Sci. Technol. 22, 053001 (2009).

2. M. Velez, J. I. Martin, J. E. Villegas, A. Hoffmann, E. M. Gonzalez, J. L. Vicent, and Ivan K. Schuller, J. Magn. Magn. Mat. 320, 2547 (2008).

3. D. Perez de Lara, F. J. Castaño, B. G. Ng, H. S. Körner, R. K. Dumas, E. M.Gonzalez, Kai Liu, C. A. Ross, Ivan K. Schuller and J. L. Vicent, Appl. Phys. Lett. 99, 182509 (2011).

4. M. J. Van Bael, K. Temst, V. V. Moshchalkov, and Y. Bruynseraede Phys. Rev.B 59, 14674 (1999).

5. M. J. Van Bael, J. Bekaert, K. Temst, L. Van Look, V.V. Moshchalkov, and Y. Bruynseraede, G. D. Howells, A. N. Grigorenko, and S. J. Bending, G. Borghs, Phys. Rev. Lett. 86, 155 (2001).

6. J. E. Villegas, E. M. Gonzalez, M. I. Montero, Ivan K. Schuller and J. L. Vicent, Phys. Rev.B 72, 064507 (2005).

7. M. Baert, V. V. Metlushko, R. Jonckheere, V. V. Moshchalkov, and Y. Bruynseraede, Phys. Rev. Lett. 74, 3269 (1995)

8. S. Avci, Z. L. Xiao, J. Hua, A. Imre, R. Divan, J. Pearson, U. Welp, W. K. Kwok, and G. W. Crabtree, Appl. Phys. Let. 97, 042511 (2010).

9. J. E. Villegas, E. M. Gonzalez, Z. Sefrioui, J. Santamaria, and J. L. Vicent , Phys. Rev. B 72, 174512 (2005).

10. J. I. Martin, M. Velez, J. Nogues and Ivan K. Schuller, Phys. Rev. Lett. 79, 1929 (1997).

11. J. I. Martin, M. Velez, A. Hoffmann, Ivan K. Schuller and J. L. Vicent, Phys. Rev. Lett. 83 1022 (1999).

12. M. Velez M, D. Jaque, J. I. Martin, M. I. Montero, Ivan K. Schuller and J. L. Vicent, Phys. Rev.B 65 104511 (2002).

13. A. V. Silhanek, L. Van Look, S. Raedts, R. Jonckheere and V. V. Moshchalkov, Phys. Rev.B 68, 214504 (2003).

14. G. Grimaldi, A. Leo, A. Nigro, A. V. Silhanek, N. Verellen, V. V. Moshchalkov, M. V. Milošević, A. Casaburi, R. Cristiano, and S. Pace, Appl. Phys. Lett. **100**, 202601 (2012).

15. A.V. Silhanek, N. Verellen, V. Metlushko, W. Gillijns, F. Gozzini, B. Ilic, V.V. Moshchalkov, Physica C **468**, 563 (2008).

16. D. Perez de Lara, F. J. Castaño, B. G. Ng, H. S. Körner, R. K. Dumas, E. M.Gonzalez, Kai Liu, C. A. Ross, Ivan K. Schuller and J. L. Vicent, Phys. Rev. B 80, 224510 (2009).

17. J. E. Davies, J. Wu, C. Leighton, and K. Liu, Phys. Rev. B, 72, 134419 (2005).

18. R. K. Dumas, C. P. Li, I. V. Roshchin, I. K. Schuller and K. Liu, Phys. Rev. B 75, 134405 (2007).

19. B. J. Kirby, S. M. Watson, J. E. Davies, G. T. Zimanyi, Kai Liu, R. D. Shull, and J. A. Borchers, J. Appl. Phys., **105**, 07C929 (2009).

20. J. Wong, P. Greene, R. K. Dumas, and K. Liu, Appl. Phys. Lett., 94, 032504 (2009).

21. M. Donahue and D. Porter, OOMMF User's Guide, Version 1.0, NISTIR6376 (1999).

22. D. J. Morgan, and J. B. Ketterson, Phys. Rev. Lett. 80, 3614 (1998).

23. D. Perez de Lara, A. Alija, E. M. Gonzalez, M. Velez, J. I. Martin, and J. L. Vicent, Phys. Rev. B 82, 174503 (2010).

**Figure captions**

Figure 1: Resistance vs perpendicular applied magnetic fields at temperature $T=0.985T_c$. Blue triangles show data obtained for the out-of plane magnetization (OP) sample, red squares for the in-plane single domain (IPSD) sample and green dots for the remanent magnetic vortex state (VS) sample. Inset shows the magnetic stray field in each configuration calculated by OOMMF simulation.

Figure 2: Resistance vs perpendicular applied magnetic fields at temperature $T=0.985T_c$ for the sample with out of plane magnetization (OP sample). Blue dots show data obtained for the ac-demagnetized state, red pointing up triangles for positive saturation remanence state and green pointing down triangles for negative saturation remanence state. Inset (a) show the sketch of the magnetic dot and the arrow shows the direction of the magnetization. Inset (b) shows the FORCs for a witness sample of the $[Pd(0.6nm)/Co(0.4nm)]_{40}$ film.

Figure 3: Resistance vs perpendicular applied magnetic fields at temperature $T=0.985T_c$ for the sample with vortex state magnetization (VS sample). Blue dots show data obtained for the disordered state with random polarity, red triangles for ordered state with aligned polarity. Inset (a) is a schematic of the vortex state with in-plane magnetization in the magnetic dot. Inset (b) shows the families of FORCs for the magnetic vortex state (VS) magnetic dot array.

Figure 4: Resistance vs perpendicular applied magnetic fields at temperature $T=0.985T_c$ for the sample with in-plane single domain magnetization (IPSD sample). Blue dots show data obtained for the demagnetized state, red triangles for positively magnetized state. Inset (a) show the sketch of the dot and the arrow shows the direction of the magnetization. Inset (b) shows the families of FORCs for the IPSD magnetic dot array.

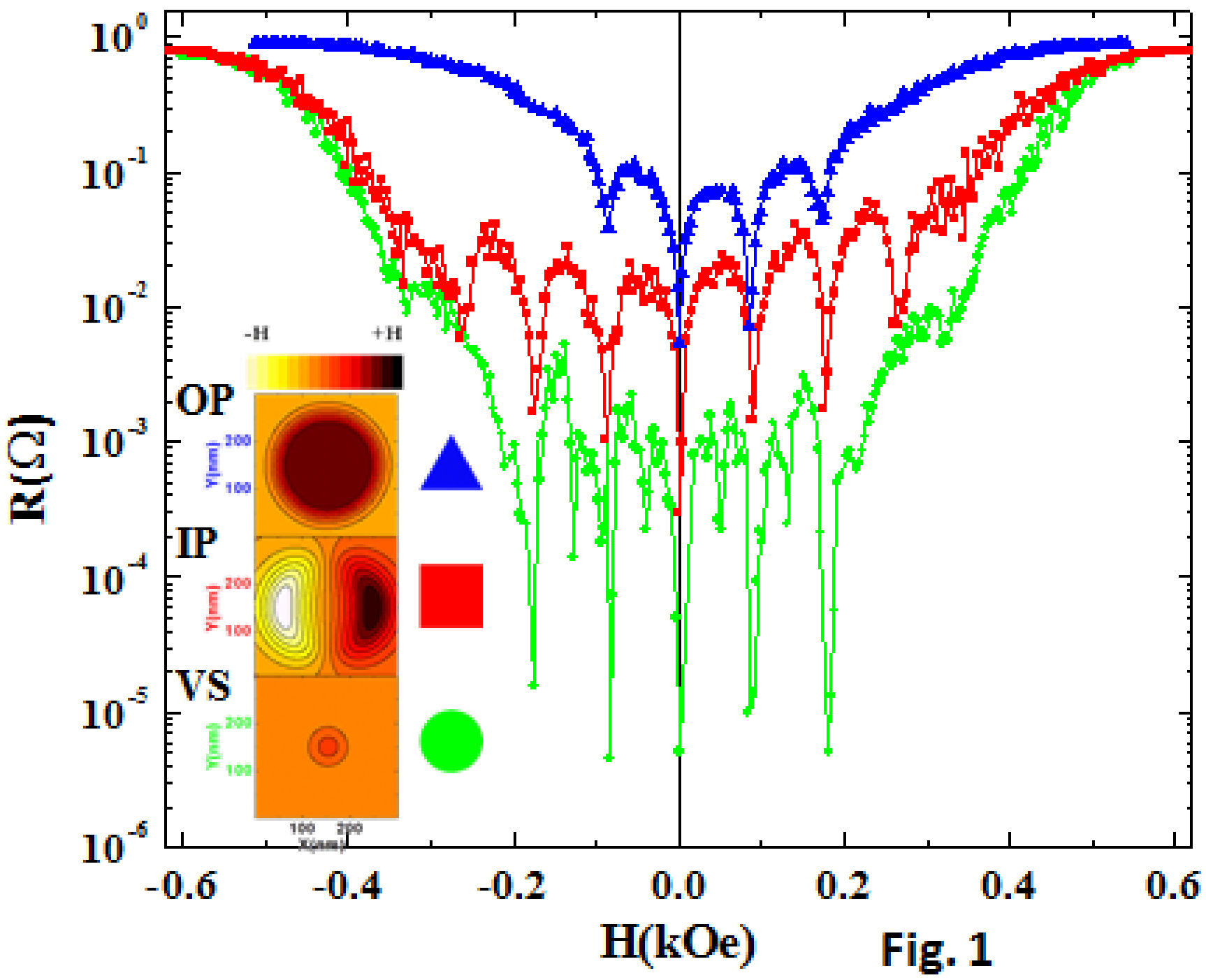

Fig. 1

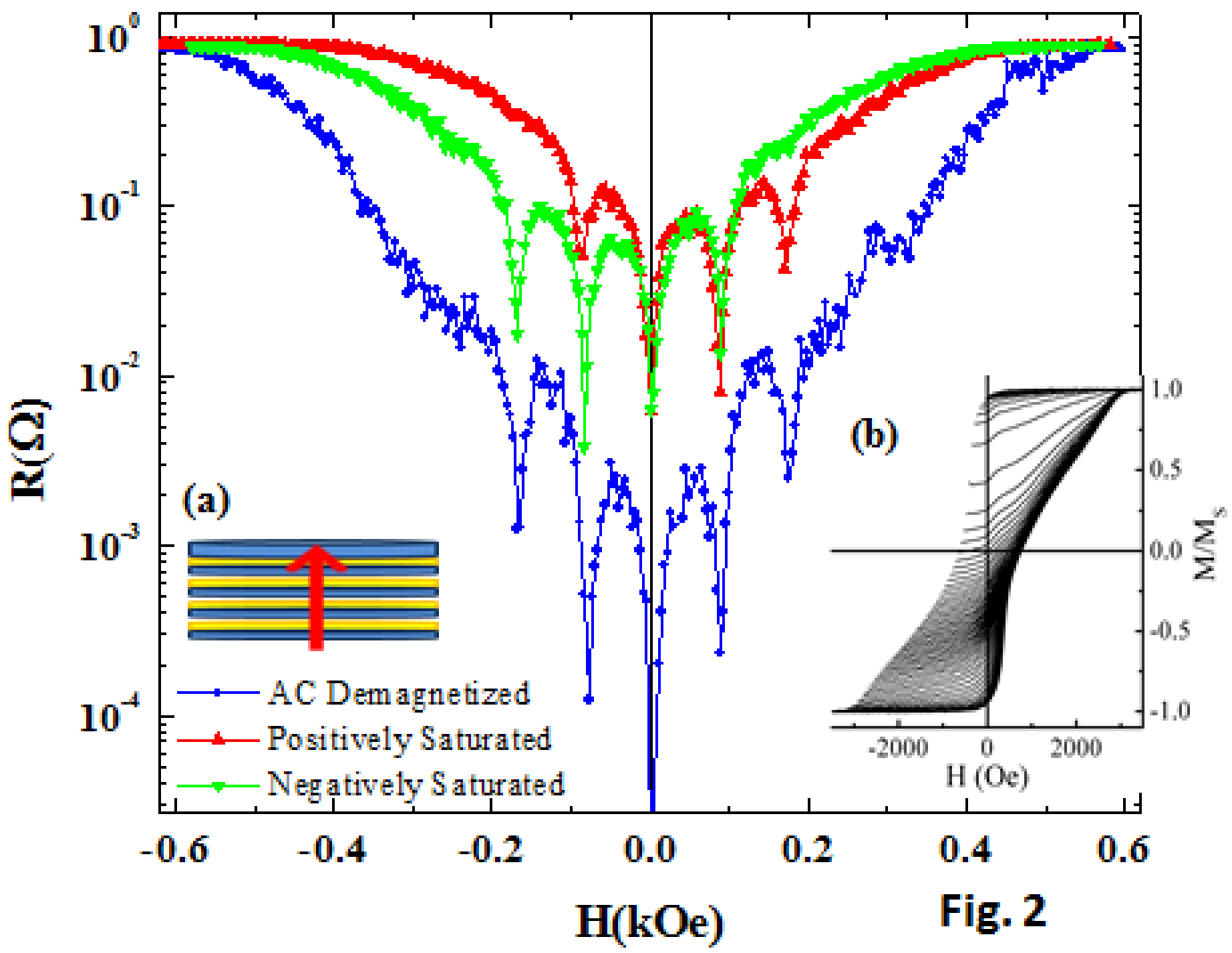
R(Ω)
H(kOe)
(a)
(b)
AC Demagnetized
Positively Saturated
Negatively Saturated
M/M_s
H (Oe)

Fig. 2

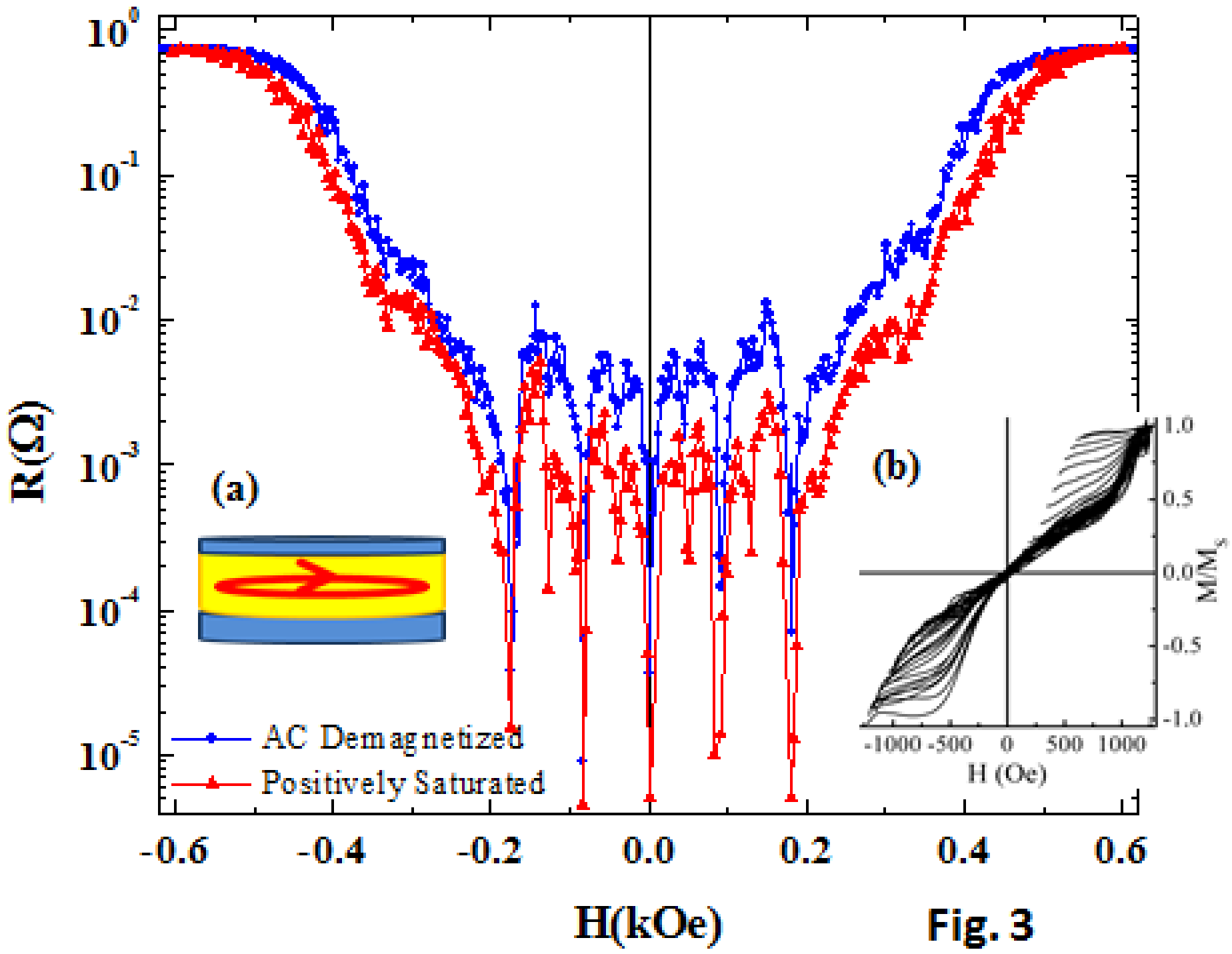

Fig. 3

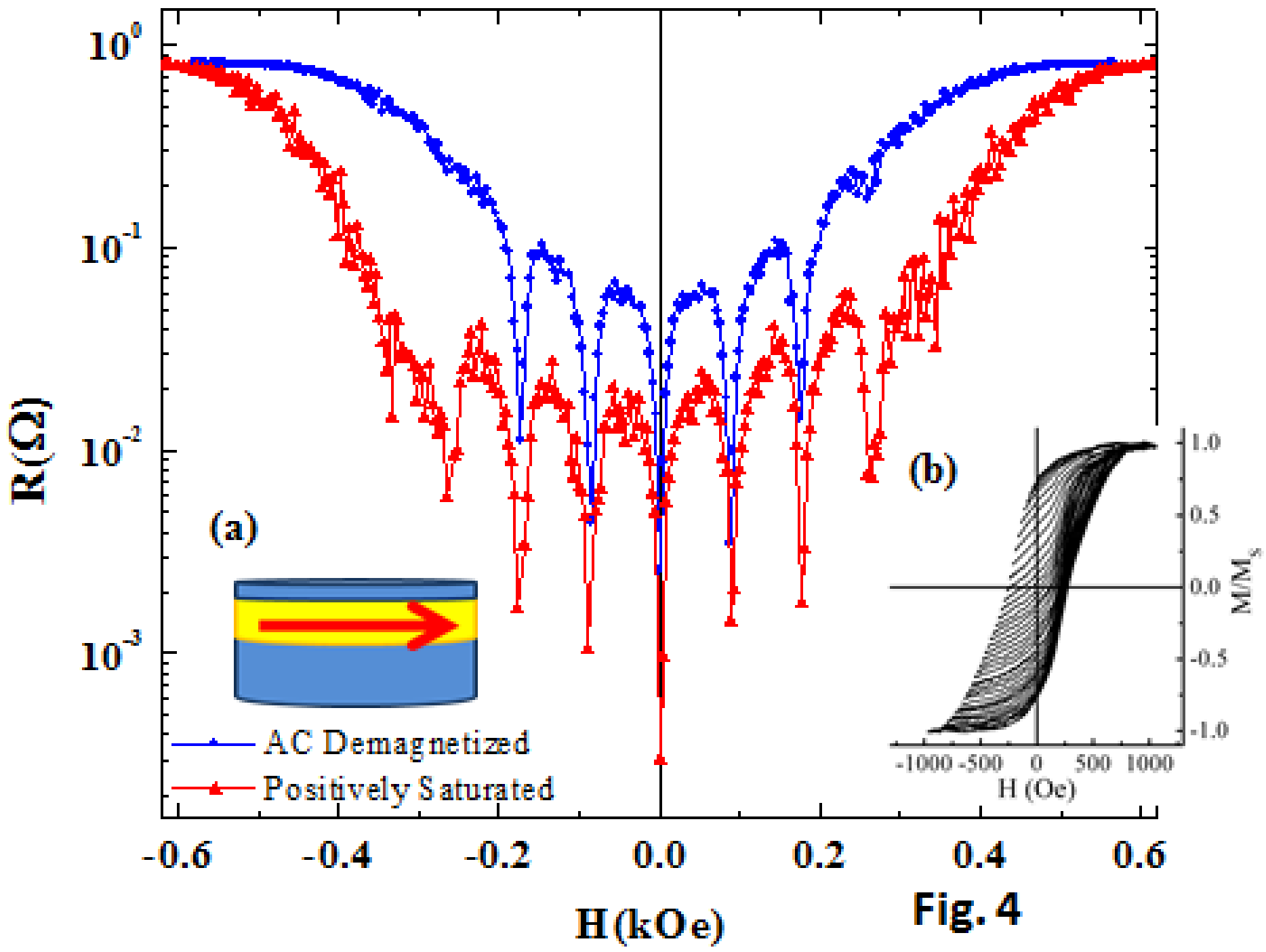

$10^0$
$10^{-1}$
$10^{-2}$
$10^{-3}$
R(Ω)
(a)
AC Demagnetized
Positively Saturated
(b)
1.0
0.5
0.0
-0.5
-1.0
$M/M_s$
-1000 -500 0 500 1000
H (Oe)
-0.6 -0.4 -0.2 0.0 0.2 0.4 0.6
H (kOe)

**Fig. 4**